\documentclass[aps,prl,twocolumn]{revtex4}
\usepackage{amsmath}
\usepackage{amsfonts}
\usepackage{amssymb}
\usepackage{amsthm}
\usepackage{graphicx}

\begin{document}

\title{Zero-field magnetization reversal of two-body Stoner particles with
dipolar interaction}
\author{Zhouzhou Sun}
\author{Alexander L\'{o}pez\footnote{Permanent address: Centro de F\'{i}sica,
Instituto Venezolano de Investigaciones
Cient\'{i}ficas, Caracas 1020-A, Venezuela}}
\author{John Schliemann}
\affiliation{Institute for Theoretical Physics, University of
Regensburg, D-93040
Regensburg, Germany}
\date{\today}

\maketitle
{\bf
Nanomagnetism has recently attracted
explosive attention, in particular, because of the enormous potential
applications in
information industry, e.g. new harddisk technology,
race-track memory\cite{Parkin},
and logic devices\cite{Allwood}.
Recent technological advances\cite{Shouheng} allow for the fabrication of
single-domain magnetic nanoparticles
(Stoner particles), whose magnetization dynamics have been extensively
studied, both experimentally and
theoretically, involving magnetic fields
\cite{Back,Schumacher,Thirion,llg1,Sun1,Sun2b} and/or by
spin-polarized
currents\cite{Tsoi,Jzsun,Myers,Kiselev,Slonczewski1,Berger,Bazaliy,Waintal,
Brataas,Stiles,Wang3}. From an industrial point of view, important issues
include lowering the
critical switching field $H_c$, and achieving short
reversal times. Here we predict a new technological perspective:
$H_c$ can be dramatically lowered (including $H_c=0$) by
appropriately engineering the
dipole-dipole
interaction (DDI) in a system of two synchronized Stoner particles.
Here, in a modified Stoner-Wohlfarth (SW) limit,
both of the above goals can be achieved.
The experimental feasibility of realizing our proposal is
illustrated on the example of cobalt nanoparticles. }


The magnetization dynamics of two Stoner particles being subject to DDI
and an external magnetic field is governed by the Landau-Lifshitz-Gilbert
(LLG) equation\cite{llg1,Landau,Gilbert},
\begin{equation}
\dot{\vec{m}}_i= -\vec{m}_i\times\vec{h}^{t}_{i}
+\alpha \vec{m}_i\times \dot{\vec{m}}_i\,.\label{LLG1}
\end{equation}
Here $\vec{m}_i=\vec{M}_i/M_s$ is the normalized magnetization vector of the
$i$-th particle, $(i=1,2)$. $M_s=|\vec{M}_i|$ is the saturation magnetization
of either particle, and $\alpha$ is the Gilbert damping coefficient.
For simplicity, we are assuming the two particles to be completely
identical in shape, volume, $\alpha$, and $M_s$.
The unit of time is set to be $(|\gamma|M_s)^{-1}$, where $\gamma$
is the gyromagnetic ratio, and the total effective field
$\vec{h}^{t}_i$ is given by
$\vec{h}^{t}_i=-\partial E/ \partial\vec{m}_i$, where
\begin{equation}
E=-\sum_{i=1,2} (k m_{i,z}^2 + \vec{m}_i \cdot \vec{h})
+\eta[\vec{m}_1 \cdot\vec{m}_2-3(\vec{m}_1\cdot\hat{n})
(\vec{m}_2\cdot \hat{n})]\,\label{energy}
\end{equation}
is the total energy per particle volume $V$ in
units of $\mu_0M_s^2$ ($\mu_0$: vacuum permeability).
Here both particles have their easy axis (EA) along the $z$-direction,
and the parameter $k$ summarizes both shape and exchange
contributions to the magnetic anisotropy.
In addition, $\vec{h}=\vec{H}/M_s$ where $\vec{H}$ is a homogeneous and static external
field. The parameter $\eta\equiv \frac{V}{4\pi d^3}$ is a
geometric factor characterizing the DDI with
$d$ being the fixed distance between the two particles
whose direction is described by the unit vector $\hat{n}$.
Here we omit the exchange interaction energy between the two particles
since it becomes important only at very small particle distances.
Moreover, in the synchronized magnetic dynamics to be investigated below,
it only contributes a constant to the energy
and will therefore not change the physical behavior.

\begin{figure}[htbp]
 \begin{center}
\includegraphics[width=5.6cm, height=5.5cm]{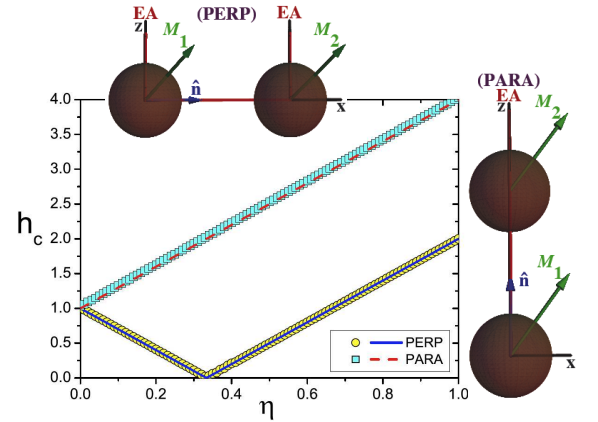}
 \end{center}
\caption{\label{fig1} The (normalized) critical switching field $h_c$ versus
the DDI strength $\eta$ for PERP and PARA configuration illustrated in the
insets. Analytical results (solid and dashed lines) are compared with
numerical findings (circles and squares).
The system parameters are  $k=0.5$ and $\alpha=0.1$. }
\end{figure}


Let us focus on two typical configurations where the connecting unit vector
$\hat{n}$ is either perpendicular or parallel to the anisotropy axes,
referred to as {\it PERP} and {\it PARA} configuration, respectively
(see insets in Fig.~\ref{fig1}). That is, introducing spherical
coordinates
$\vec{m}_i=(\sin\theta_i\cos\phi_i, \sin\theta_i\sin\phi_i, \cos\theta_i)$,
$\hat{n}=(\sin\theta_n\cos\phi_n, \sin\theta_n\sin\phi_n, \cos\theta_n)$,
we have $\theta_n=\pi/2$ for PERP configuration and $\theta_n=0$ for PARA
(or $\theta_n=\pi$ since DDI is invariant under $\hat{n} \mapsto -\hat{n}$).
Moreover, without loss of generality, we can let $\phi_n=0$ by choosing the
$x$-axis along the connecting line of the two particles.
Furthermore, we concentrate on the synchronized magnetic dynamics of the two Stoner
particles, where both magnetization vectors remain in parallel throughout
the motion, $\theta_1=\theta_2=\theta$, $\phi_1=\phi_2=\phi$.
Thus, the two particles behave like a single entity,
and this two-body Stoner particle can be regarded as a computer
information bit. We have verified by numerical
simulations (see discussion below) that this dynamical regime is stable
against perturbations.
For this synchronized motion mode, the
nonlinear coupled LLG equations (\ref{LLG1}) read in spherical coordinates
\begin{align}
&\dot{\theta}+\alpha \sin\theta\dot{\phi}=-3\eta\cos\psi\sin
\theta_n\sin\phi\,,\nonumber\\
&\alpha\dot{\theta}-\sin\theta\dot{\phi}=h\sin\theta-k\sin2\theta
+\frac{3\eta}{2}\frac{\partial\cos^2\psi}{\partial\theta}\,,\label{llg2}
\end{align}
where $\psi$ is the angle between $\vec{m}$ and
 $\hat{n}$,
i.e. $\cos\psi=\cos\theta\cos\theta_n +\sin\theta\sin\theta_n\cos\phi$.
Here we have put $\vec{h}=-h\hat{z}$ (antiparallel) along the easy axis,
which is the usual field configuration for reversing a magnetic bit.
The above equations are the starting point of our numerical calculations
to be discussed below.

However, in order to analytically explore the SW limit for magnetic
reversal\cite{Stoner,llg1},
we assume the external field to lie in the plane
spanned by
the easy axis and the interparticle direction,
$\vec{h}=h_z\hat{z}+h_x\hat{x}$.  The energy takes the form
$E=-2k\cos^2\theta -2 h_z\cos\theta -2h_x\sin\theta +\eta(1-3\cos^2\psi)$
with $\psi=\theta-\theta_n$.
The SW limit occurs at the inflection of the energy as a function
of $\theta$,
i.e. $\partial E/\partial \theta=\partial^2 E/\partial \theta^2=0$.
An elementary calculation translates this condition into
\begin{equation}
\left(\frac{h_x}{2k\mp 3\eta}\right)^{2/3}
+\left(\frac{h_z}{2k\mp 3\eta}\right)^{2/3}=1,
\end{equation}
where the minus(plus) sign corresponds to the PERP(PARA) configuration,
respectively. Note that in the absence of DDI, $\eta=0$,
the above equation just recovers the usual SW limit for a single Stoner
particle.
As a result, the critical switching field $h_c$
(applied antiparallel to the easy axis) is given by
\begin{equation}
h^{PERP}_c=|2k-3\eta|, \qquad  h^{PARA}_c=2k+3\eta. \label{hc1}
\end{equation}
This analytical solution \eqref{hc1} is shown by the solid and
dashed lines in Fig.~\ref{fig1}.

Moreover, the above findings imply the remarkable observation
that in the PERP configuration there exists a critical DDI strength
$\eta_c=2k/3$ such that {\it the critical switching field vanishes!}
($h_c=0$).\\
\noindent The zero-field condition is achieved for interparticle distances
given by
\begin{equation}
d_c=\left(\frac{3\mu_0M_s^2V}{8\pi K}\right)^{1/3},
\end{equation}
where $K=k\mu_0M_s^2$ is the standard anisotropy coefficient.
Remarkably, the above condition is independent of the damping, and
its physical contents can be illustrated in terms of energy landscape.
In the PERP configuration, the energy in the absence of an external
field is $E=-2\eta+3(\eta-\eta_c)\cos^2\theta$. Thus, for $\eta=\eta_c$,
any angle $\theta$ represents an equilibrium position such that
an arbitrary small field is sufficient to move the magnetization
along the field axis, implying the possibility of zero-field reversal.
Moreover, for $\eta>\eta_c$, the zero-field ground state is given by
$\theta=\pi/2$, i.e. the synchronized magnetization points along the
interparticle axis, while for $\eta<\eta_c$, the ground state magnetization is
along the easy axis, $\theta=0, \pi$.


In order to complement and quantitatively support our previous discussion,
we have performed numerical simulations of the LLG equations~\eqref{llg2}
using the 4th-order Runge-Kutta scheme.
We consider a range of the DDI parameter of $0\leq \eta \leq 1$.
The case, $\eta=0$ corresponds to the limit $d\rightarrow\infty$, i.e.
the two nanoparticles being infinitely apart.
Large $\eta$ can be realized by fabricating magnetic particles
of ellipsoidal shape allowing for a closer proximity.
Throughout the numerical results shown here, we use a damping parameter of
$\alpha=0.1$. In Fig.~\ref{fig1} we compare simulation results
for the critical switching field with the analytical formulae \eqref{hc1}, both
findings being in excellent agreement.
The critical switching field in the PARA configuration is always higher
than the value without DDI.
Thus, only the PERP configuration will be useful for possible technology
applications.

\begin{figure}[htbp]
 \begin{center}
\includegraphics[width=7.5cm, height=5.5cm]{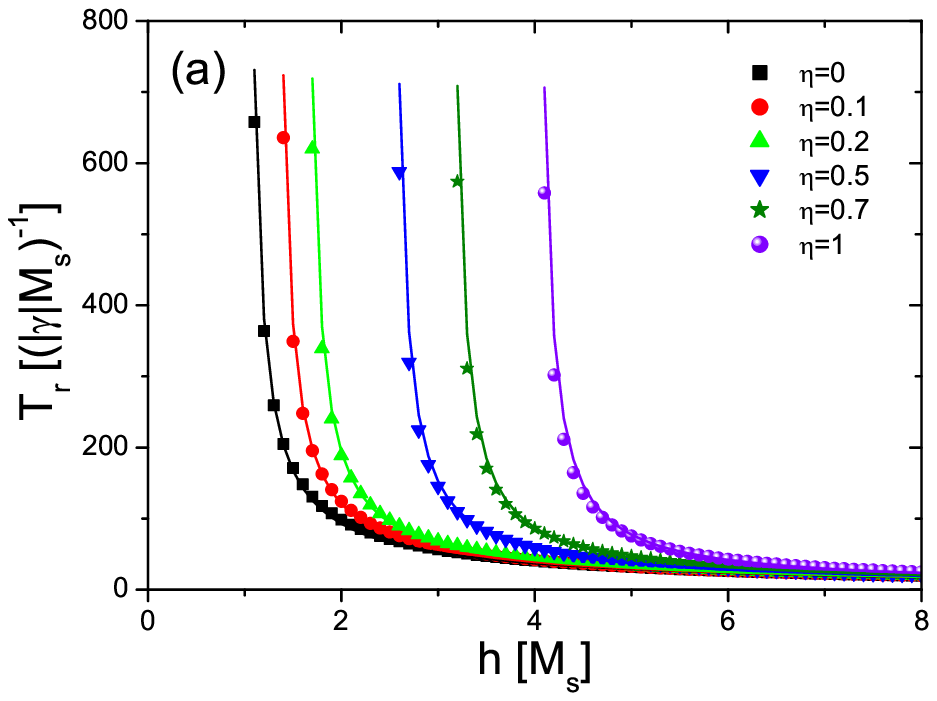}
\includegraphics[width=7.5cm, height=5.5cm]{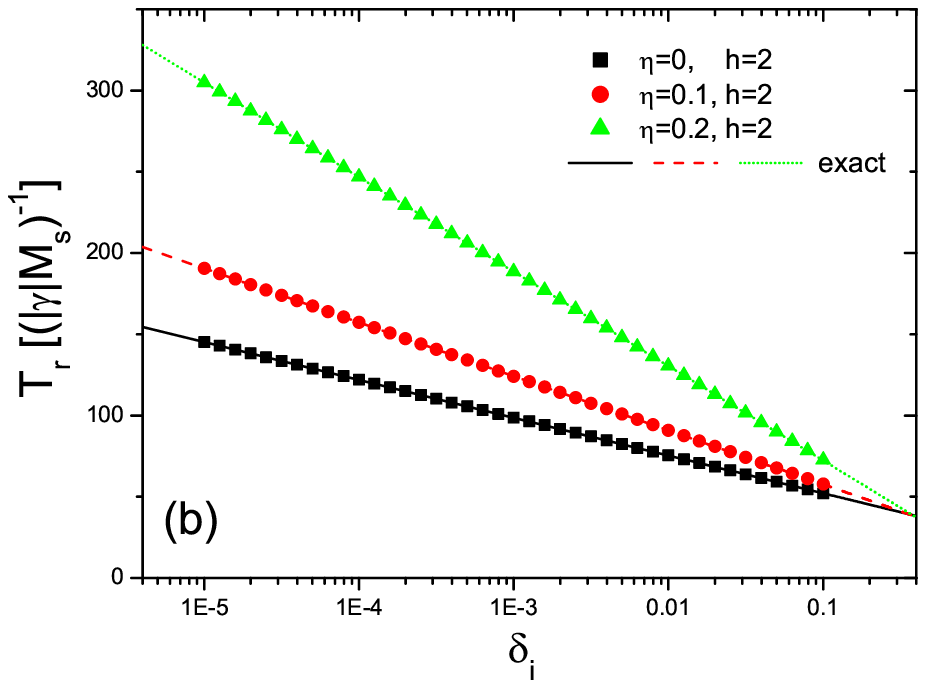}
 \end{center}
\caption{\label{fig2} PARA configuration: (a) Reversal time $T_r$
versus field magnitude $h$ at different DDI strength $\eta$.
The simulation data (symbols) is fitted according to the approximate
expression $T_r \simeq \frac{2\lambda h(-\ln\delta_i)}{h^2-h_0^2}$
for $\delta_i=\delta_f=0.001$,  and $k=0.5$, $\alpha=0.1$ (cf. Fig~\ref{fig1}).
(b) Dependence of $T_r$ on the initial condition $\delta_i$: Simulation data
along with exact analytical results according to Eq.~(\ref{Tr}).
The final state is fixed to be $\delta_f=0.001$.}
\end{figure}

\begin{figure}[htbp]
 \begin{center}
\includegraphics[width=7.5cm, height=5.5cm]{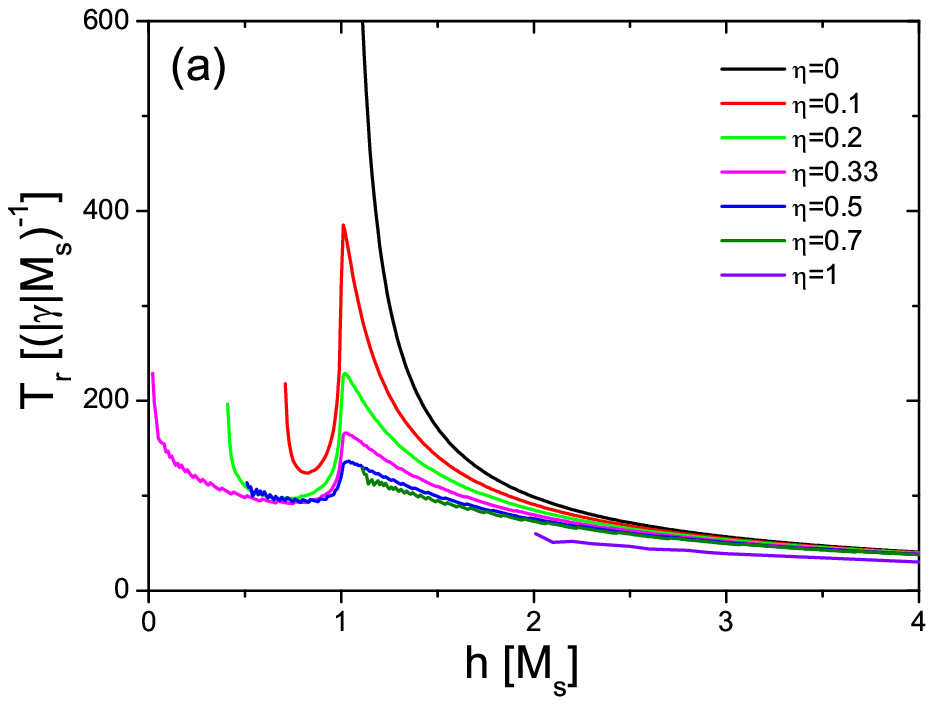}
\includegraphics[width=7.5cm, height=5.5cm]{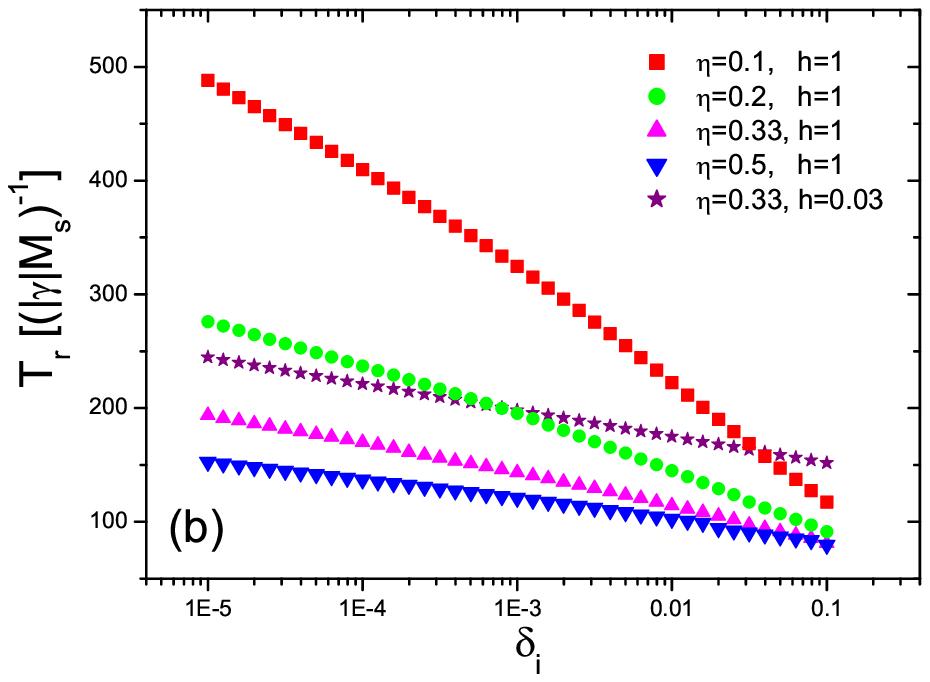}
 \end{center}
\caption{\label{fig3} PERP configuration: (a)  Reversal time $T_r$
obtained from numerical simulations
versus field magnitude $h$ at different DDI strength $\eta$.
(b) $T_r$ as a function the initial condition $\delta_i$. All other
parameters are the same as Fig.~\ref{fig2}.}
\end{figure}

Let us now turn to the reversal time $T_r$, i.e. the duration of the
reversal process of the magnetization direction changing the angle
$\theta$ from $0$ to $\pi$.
To avoid the metastable points  $\theta=0,\pi$,
we introduce two small deviations
$\delta_i,\delta_f$, by defining $\theta_i=\delta_i$
(i.e. $m_{z}\approx 1$) and $\theta_f=\pi-\delta_f$
(i.e. $m_{z}\approx -1$), which leads to a finite
reversal time in our simulations.
In the PARA configuration, one can derive a closed analytical expression
for this quantity,
\begin{equation}
T_r =-\lambda\left[\frac{\ln\delta_i}{h-h_0}+\frac{\ln\delta_f}{h+h_0}
+\frac{h_0 \ln(\frac{h+h_0}{h-h_0})-h\ln 4}{h^2-h_0^2}\right]\,,
\label{Tr}
\end{equation}
where $h_0=2k+3\eta$ and $\lambda=(1+\alpha^2)/\alpha$. Note that the case
$\eta=0$ also includes the PERP configuration since both configurations are
indistinguishable here.

In the upper panel (a)
of  Fig.~\ref{fig2} we show simulation
results for $T_r$ in the PARA configuration
for again $k=0.5$, $\delta_i=\delta_f=0.001$, and
various $\eta$. The simulation data is well described by the approximate
expression $T_r \simeq \frac{2\lambda h(-\ln\delta_i)}{h^2-h_0^2}$
valid for small $\delta_i=\delta_f \ll 1$.
As a result, in the PARA configuration
the reversal time increases with increasing strength of the dipolar interaction.
The sensitivity of the data
to the parameters $\delta_i$ (where $\delta_f$ is fixed or vice versa)
is illustrated in the bottom panel (b)
of Fig.~\ref{fig2}: In accordance with Eq.~(\ref{Tr}), the reversal time
depends only logarithmically on the quantity and does
not change its order of magnitude while $\delta_i$ is changing
over several powers of ten. Thus our results are not qualitatively affected
by our choice of the condition $\delta_i=\delta_f=0.001$.

A similar result regarding the sensitivity to initial conditions
is obtained in the PERP configuration. Here an analytical result
comparable to Eq.~(\ref{Tr}) does not seem to be achievable. However,
the simulation data is shown in Fig.~\ref{fig3}(b) that $T_r$
again depends only logarithmically on $\delta_i$ and is
therefore similarly insensitive to the initial condition as in the PARA case.

The dependence of the reversal time
on the dipolar parameter $\eta$ shown in Fig.~\ref{fig3}(a),
however, is strikingly different from the
PARA configuration: Here $T_r$ clearly decreases with increasing $\eta$,
especially in the regime of low fields, which also strongly favors
the two-body bit implementation proposed in this letter.

\begin{figure}[htbp]
 \begin{center}
\includegraphics[width=5.5cm, height=5.1cm]{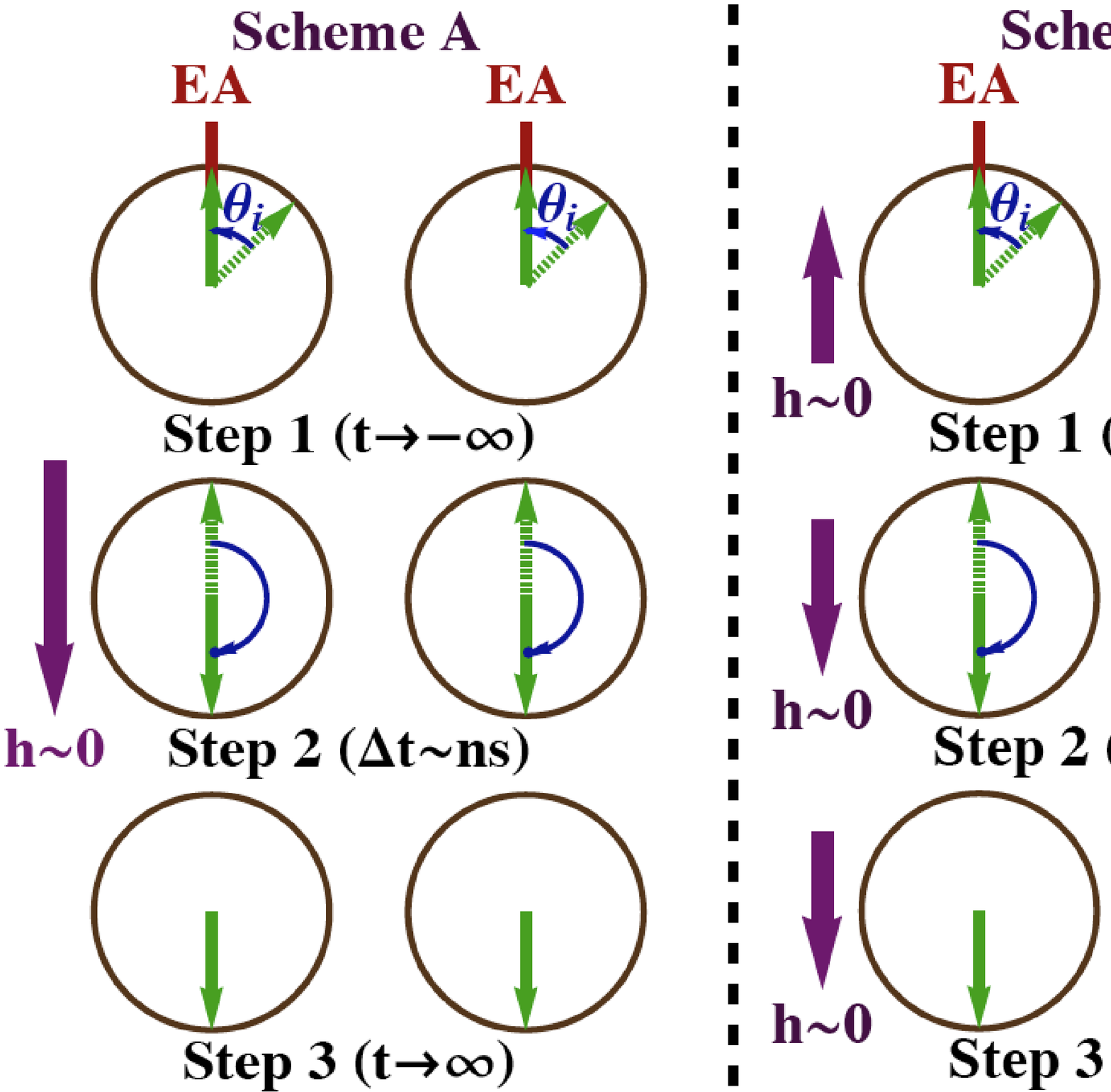}
\includegraphics[width=8.5cm, height=5.5cm]{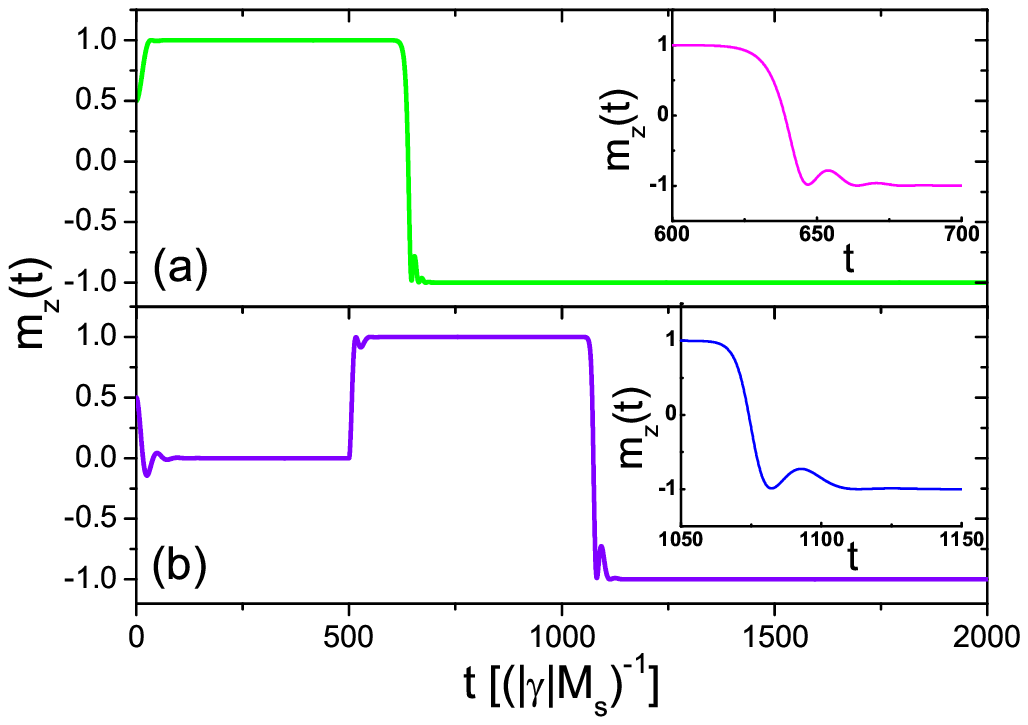}
\includegraphics[width=8.5cm, height=5.5cm]{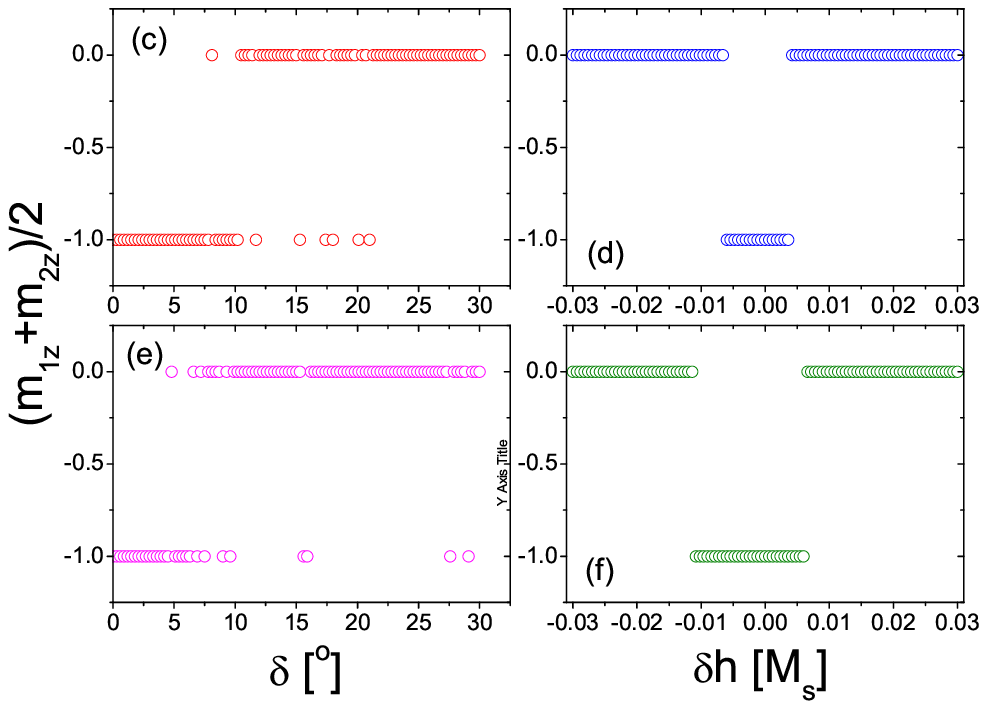}
 \end{center}
\caption{\label{fig4} Upper panels: Two schemes of magnetization reversals
of the two-body Stoner particle (see text).\\
Middle panels: $m_z(t)$ versus time at (a) $\eta=0.33<\eta_c=1/3$,
(b) $\eta=0.34>\eta_c=1/3$. In (a), an antiparallel field $h=0.03$
is applied only during the time interval $500<t<1000$ with a reversal time
of $T_r\approx 198(|\gamma|M_s)^{-1}$; In (b), a parallel field $h=-0.03$ is applied
during $500<t<1000$ and then reversed to $h=0.03$, leading
to a reversal time as $T_r\approx 155(|\gamma|M_s)^{-1}$. The insets show the
reversal process on a smaller scale.\\
Bottom panels (c) \& (e): Average stable magnetization $(m_{1z}+m_{2z})/2$ versus
the deviation angle $\delta=\theta_{2}-\theta_{1}$ (in degrees)
between the initial magnetization directions
in the two schemes A and B, respectively.
Panels (d) \& (f): Analogous data as a function of
the deviation field $\delta h=h_2-h_{1}$ antiparallel to the
easy axis (see text).}
\end{figure}

Thus, around the critical DDI strength $\eta_c$ we find not only nearly a zero
critical switching field but also a substantially shorter reversal time.
This result promises attractive future applications for computer technology,
such as fast read/write hard disk or magnetic RAM.
Let us discuss two schematic setups for experimentally realizing
the zero-field switching mechanism.
The first scheme A is illustrated in the upper-left
panel of Fig.~\ref{fig4}, along with a numerical simulation in
panel (a). Here we chose again $k=0.5$, and the DDI parameter
is $\eta=0.32<\eta_c=1/3$. Thus, the magnetization in the zero-field ground
state points along the easy axis ($z$-axis).
Implementation steps of scheme A are demonstrated as follows:
{\it Step 1} - First we choose a casual initial condition,
for instance $m_{1z}(0)=m_{2z}(0)=0.5$,
to mimic the relaxation of the system to its ground state
under zero-field during $0<t<500$;
{\it Step 2} - Then we apply a tiny antiparallel field
$h=0.03$ during $500<t<1000$ which drives the magnetization
reversal process of the two-body Stoner particle along the field
direction. The reversal time is found to be
$T_r\approx 198 (|\gamma|M_s)^{-1}$,  and the inset shows the reversal process
on a smaller scale; {\it Step 3} - Finally we quench the field and obtain
the new stable magnetic state.

The second scheme B is sketched in the upper-right panel of
Fig.~\ref{fig4} along with a numerical simulation in middle panel (b).
Here $\eta=0.34>\eta_c=1/3$ for $k=0.5$, i.e. the zero-field ground state
is along the hard axis ($x$-direction), and a field along the easy axis
is permanently required to preserve the magnetization state (information)\cite{note1}.
However, our results show that such a field
can be very small and actually close to zero. Thus, it is not implausible to
generate such a field as the Oersted field of tiny switchable currents.
The reversal process can be implemented as follows: {\it Step 1} -
We again choose the initial condition $m_{1z}(0)=m_{2z}(0)=0.5$, and the system
relaxes to its ground state ($x$-axis) during
$0<t<500$. Then we apply a tiny field $h=-0.03$
(note our definition $\vec{h}=-h\hat{z}$) to preserve the magnetic
state $m_z=1$ during $500<t<1000$;
{\it Step 2} - After
$t > 1000$ the field is reversed to $h=0.03$, and the two-body  particle
reverses its magnetization during a reversal time as
$T_r\approx 155 (|\gamma|M_s)^{-1}$ (see also inset); {\it Step 3} -
The reverse field is applied permanently to preserve the bit information.


We now address the stability of the synchronized dynamics of the
two-body particle as studied so far. Let us first investigate small
deviations of the initial magnetization directions of the two subparticles
being otherwise still identical.
In detail, we fixed the initial direction
(at $t=500$ in reversal scheme A, and at $t=1000$ for scheme B)
of one subparticle to be close to the $z$-axis, i.e. $\theta_1=0.001$,
and changed the other initial direction
to $\theta_2=0.001+\delta$.
The result is shown in Fig.~\ref{fig4}(c) and (e): For a finite range of
deviations, $\delta < 7.8^{\circ}$ ($\delta < 4.5^{\circ}$)
in scheme A(B), the synchronized mode remains stable, while for substantially
larger $\delta$, the average magnetization mostly reaches zero.

Let us now consider the case where the two particles differ slightly in
anisotropy $K_i$, volume $V_i$, saturation magnetization $M_{s,i}$ ($i=1,2$).
Here the zero-field mechanism still occurs at
$\eta_c=(k_1+k_2)/3=\frac{2(K_1V_1+K_2V_2)}{3\mu_0M_{s1}M_{s2}(V_1+V_2)}$.
However, the effective field experienced by each particle will also be
slightly different.
To numerically check the stability under such
different external fields,
we fixed the field on one subparticle to
$h_1=0.03$ and changed the field on the other subparticle as
$h_2=0.03+\delta h$. The numerical results are given in
Fig~\ref{fig4}(d) and (f) for reversal schemes A and B, respectively, and
demonstrate again a finite range of stability against deviations from
the case of strictly identical particles.

To give a concrete and practical example of our findings, let us discuss
the case of cobalt (Co) particles.
The standard data is $M_s=1400$kA/m, uniaxial strength $K=10^5$J/m$^3$,
$\alpha=0.1$\cite{Back}. Thus $k=K/(\mu_0M_s^2)= 0.04$ such that
$\eta_c=0.027$. For two spherical particles with radius $r$,
i.e. $\eta=r^3/(3d^3)$, the critical DDI strength is reached at
$d_c=2.3r$. The critical switching field without DDI (SW limit) is
$H_{SW}=2K/(\mu_0M_s)= 1400$Oe. In the presence of DDI, and considering deviations
$\Delta d$ from $d_c$, one can express the critical switching field as
$H_c/H_{SW}=3|\Delta d|/d_c$. Thus, in order to drastically reduce the
switching field by taking advantage of our proposal, one has to engineer the
interparticle distance on a scale of $d_c=2.3r$ which is typically a few hundred
nanometers. In the case of Co, the time unit is $(|\gamma|M_s)^{-1}\ =3.23$ps
rendering the reversal times in the schemes A and B to be
 $T_r\approx0.64$ns and $T_r\approx0.5$ns, respectively, which are much shorter than
in the conventional setup.

Another important issue concerns thermal fluctuations. For $\eta<\eta_c$ the
ground state ($\theta=0,\pi$) is stable if the energy barrier in presence of
DDI is large compared to the thermal energy
$3(\Delta \eta)\mu_0M_s^2 V \gg k_B T$ ($k_B$: Boltzmann's constant),
which translates for  $T=400$K and the above  parameters for Co into
$(\Delta \eta) r^3 \gg 0.18{\rm nm}^{3}$. On the other hand, the energy
scale of the applied field should also be large compared to
thermal effects, $2\mu_0 M_s H V \gg k_B T$, which, under the same conditions,
reduces to $H\gg 4700/r^3\,{\rm Oe}\,{\rm nm}^{3}$. Thus, considering a typical
particle radius of $r=100$nm, i.e. $d_c=230$nm, and $H_c/H_{SW}=0.03$,
i.e. $H_c=42$Oe, both conditions are easily satisfied,
$(\Delta \eta)r^3=810{\rm nm}^{3} \gg 0.18{\rm nm}^{3}$ and
$H_c=42$Oe $\gg 0.0047$Oe.

Finally, we would like to remark that our general results
regarding the influence of dipolar interaction on the critical
switching field of two-body Stoner particles are consistent with
recent experimental and micromagnetic simulation results showing that the
coercive field for an array of nanowires can be lower than for a single nanowire\cite{Nielsch, Hertel}.

{{\it Acknowledgments}--} We thank Christian Back for useful discussions.
This work has been supported by the Alexander von
Humboldt Foundation (ZZS), by DAAD-FUNDAYACUCHO (AL), and by
Deutsche Forschungsgemeinschaft via SFB 689.



\begin{thebibliography}{}
\bibitem{note1} One might also consider a reversal process
along the hard axis $m_x=1$ to $m_x=-1$
by applying a field along -$x$ direction. However, numerical simulations show
that the synchronized motion mode becomes unstable for such an initial
condition. This is different from the reversal process in scheme B
(see text).
\end{thebibliography}
\end{document}